\documentclass{PoS}

\PoS{PoS(LAT2005)039}

\title{Disconnected Loops with Twisted Mass Lattice QCD }

\ShortTitle{Disconnected Loops with Twisted Mass Lattice QCD }

\author{\speaker{Walter Wilcox} and Dean Darnell\\

        Baylor University\\ Department of Physics, Waco, TX, USA 76798-7316\\

        E-mails: \email{walter\_wilcox@baylor.edu},\,\,\email{dean\_darnell@baylor.edu}}

\author{Ron Morgan\\

        Baylor University\\ Department of Mathematics, Waco, TX, USA 76798-7328\\

        E-mail: \email{ron\_morgan@baylor.edu}}

\author{Randy Lewis\\

        University of Regina\\ Department of Physics, Regina, SK, Canada S4S-0A2\\

        E-mail: \email{randy.lewis@uregina.ca}}

\abstract{We give a general introduction and discussion of the
issues involved in using the twisted mass formulation of lattice
fermions in the context of disconnected loop calculations,
including a short orientation on the present experimental
situation for nucleon strange quark form factors. A prototype
calculation of the disconnected part of the nucleon scalar form
factor is described.}

\FullConference{XXIIIrd International Symposium on Lattice Field Theory\\

                 25-30 July 2005\\

                 Trinity College, Dublin, Ireland}

\begin{document}

\section{Disconnected tmLQCD Calculations}
Twisted mass fermions have great potential in extending lattice
QCD calculations to smaller quark masses. It is essentially a
chiral-flavor rotation of the mass term in the Wilson action for a
quark doublet. It does not suffer from the quenched \lq\lq
exceptional configuration" problem because of the suppression of
unphysical small quark mass eigenvalues\cite{zero}. In addition,
$O(a)$ improvement is often automatic in many
quantities\cite{half}. Although twisted mass fermions are
beginning to be used in phenomenological studies\cite{one}, their
use in the context of lattice disconnected loop calculations is
still in the initial stages. Our goal is to make realistic
calculations of experimental observables, especially nucleon
strange form factors. Previously, we have used Wilson fermions in
our disconnected calculations\cite{two}. The improved behavior of
small quark eigenvalues in the twisted formulation changes our
previous computational strategy considerably. We will discuss our
computational approach below, after a brief review of the current
experimental situation on strange form factors.

\section{Experiments on Strange Quark Form Factors}

The current experimental situation on low momentum transfer
measurements of nucleon strange form factors from
HAPPEX\cite{happex}, A4\cite{a4}, SAMPLE\cite{sample} is nicely
summarized in Fig.~1 below, taken from the second paper in
Ref.\cite{happex}. (Additional experimental measurements are also
underway\cite{g0}.) It shows a simultaneous plot of the bands in
$G_E(q^2$), $G_M(q^2$) picked out by various linear combinations
measured. These measurements are deduced from parity-violating
asymmetry experiments done in elastic electron-proton scattering.
The $95\%$ confidence region of these 4 experiments is indicated
by the oval, giving small $G_E$, positive $G_M$ values. Two
lattice results are indicated on the figure, listed as references
21 and 22. The reference 21 result\cite{two} came from fitting our
quenched lattice data with two chiral perturbation theory models;
a \lq\lq no $\eta'$" model, and a \lq\lq maximal $\eta'$" model.
The \lq\lq no $\eta'$" model is more realistic, and predicts
small, positive $G_E$, $G_M$ values, in apparent agreement with
experiment. Although our lattice results agree with experiment, it
is clear that we must simulate at smaller quark masses to connect
more confidently to the chiral models. Thus, we turn to the
twisted mass formulation, which promises a deeper penetration of
the chiral region.

\begin{figure}
\begin{center}
\includegraphics[width=3.5in]{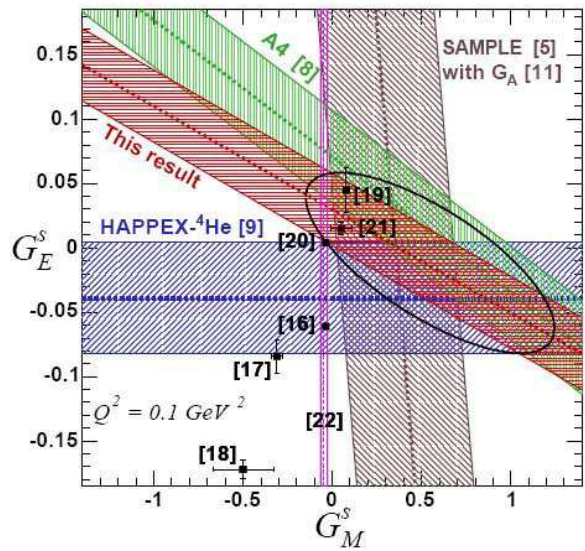}
\caption{The experimental situation for the simultaneous
measurement of the strange electric and magnetic form factors at
small four-momentum transfer. Figure taken from the second paper
in Ref.~5} \label{happexresult}
\end{center}
\end{figure}

\section{Twisted Mass Eigenvalues}
An illustrative eigenvalue spectrum, where we zoom in on the
eigenvalues near the origin, on a small $8^3\times 16$ lattice
with $\beta=6.0$ is shown in Fig.~2. These are actually harmonic
Ritz values of the eigenvalue spectrum of the preconditioned
even/odd matrix, run at $\kappa=0.1567, \mu=0.03$, which
corresponds to maximal twist as determined in
Ref.\cite{eventually}. The harmonic Ritz values are approximate
eigenvalues. The spectrum has essentially no small eigenvalues.
This is in the context of the spectrum of harmonic Ritz values (a
total of 140 shown) seen in Fig.~3 extending from -25 to 25 on the
imaginary axis, and from 0 to 50 on the real axis. Also, one sees
a symmetric pairing of up/down quark eigenvalues as one reflects
across the real axis. We see these same phenomena occuring on
larger lattices.

\begin{figure}
\begin{center}
\includegraphics[width=3.5in]{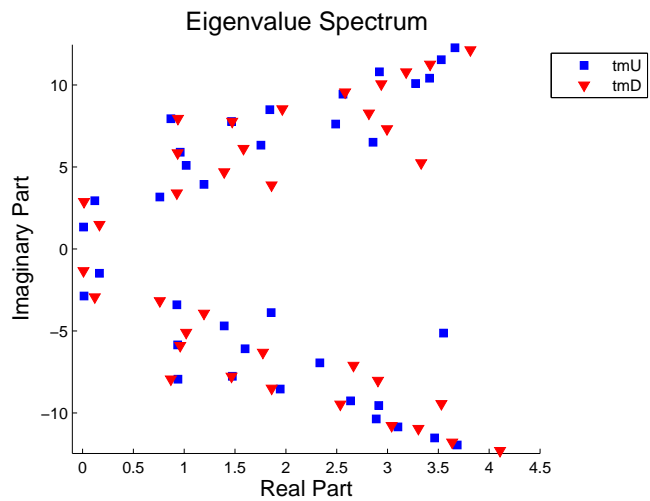}
\caption{A selection near the origin of the even/odd
preconditioned twisted mass quark eigenvalues of a small
$8^3\times 16$ lattice.} \label{dublinevals}
\end{center}
\end{figure}

\section{tmLQCD and Perturbative Subtraction}

The basic method we use to extract the subtle signals in the
lattice simulation of disconnected quantities is stochastic noise
projection. That is, we invert the tmWilson matrix with Z(2) noise
vectors placed at every color, Dirac, and space-time position.
Within this context, one of the crucial techniques we previously
used in our Wilson disconnected diagram evaluations was
perturbative subtraction. The perturbative evaluation of Wilson
matrix elements mimics the iterative numerical evaluation, and
allows a subtraction of the noise from $Z(2)$ source noise
vectors. This allows a better signal to be extracted for a given
number of input noise vectors, and can be used as long as one
restores the perturbative signal extracted along with the noise.
We tested perturbative subtraction on tmLQCD on a $20^3\times 32$
lattice with $\kappa=0.1567, \mu=0.03$ (about the strange quark
mass). Fig.~4 shows the zero momentum time sliced disconnected
signal for the imaginary part of the third component of the
conserved vector current, $J_3(t)$, for the first 16 time
positions in the lattice. We found that the method has very little
effect on the extracted signal. We ran 50 noise loops for this
quantity on a single configuration. The
Liverpool/M\"unster/Zeuthen/Hamburg/Berlin group claims a better
result with Gaussian noise\cite{four}. We will soon explore this
option as well.

\begin{figure}
\begin{center}
\includegraphics[width=3.5in]{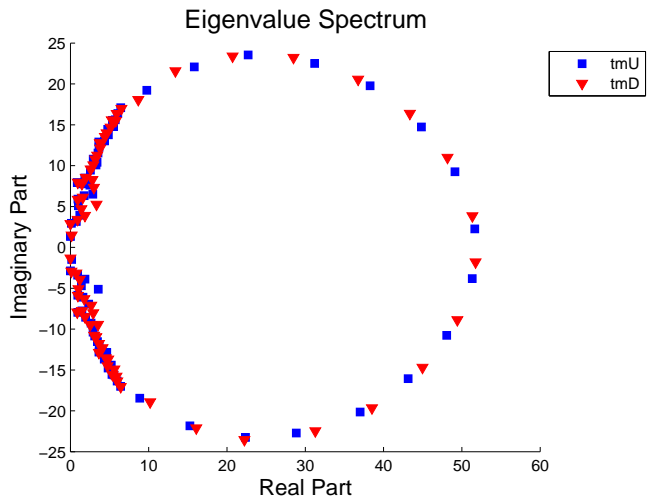}
\caption{An illustration of the full set of even/odd
preconditioned twisted mass quark eigenvalues on a small
$8^3\times 16$ lattice.} \label{fullspec}
\end{center}
\end{figure}

\begin{figure}
\begin{center}
\includegraphics[width=3.5in]{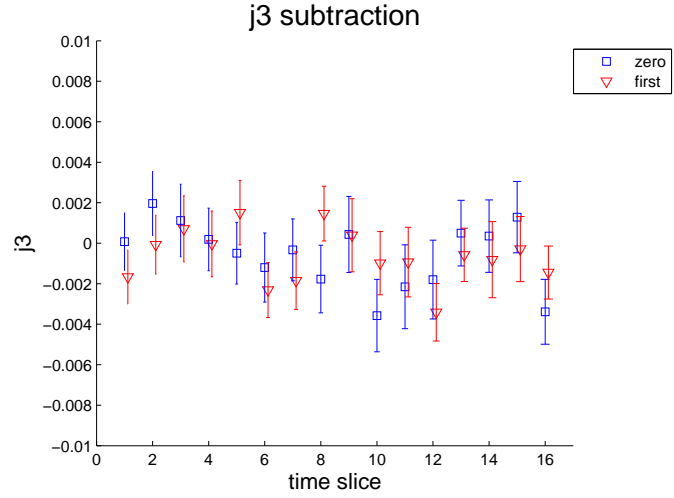}
\caption{Illustration of the effect of perturbative subtraction on
the real part of the $J_3$ twisted mass current loops for time
positions 1 through 16 on a $20^3\times 32$ lattice. "Zero" means
without subtraction, "first" means subtracted to order $\kappa^4$
in the perturbative series. Corresponding data points displaced
slightly.} \label{j3current}
\end{center}
\end{figure}

\begin{figure}
\begin{center}
\includegraphics[width=3.5in]{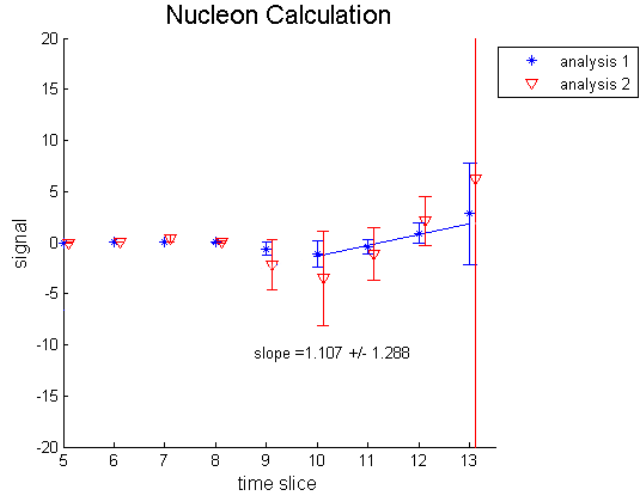}
\caption{The blue bursts and associated error bars represent an
analysis using Eq.~(5.3) on 135 configurations. The red triangles
and error bars are from a similar analysis, using a loop
background of 20 time steps, corresponding to a fixed time upper
limit in Eq.~(5.3).} \label{comparekt}
\end{center}
\end{figure}

\section{Prototype Scalar Disconnected Calculation}

In order to investigate the issues arising in extracting
disconnected signals in this new context, we have performed a
small simulation on a $20^3\times 32$ lattice, at $\kappa=0.1567,
\mu=0.03$, with twisted mass fermions. We use unimproved Wilson
gauge fields and periodic spatial, Dirichlet time boundary
conditions on the fermion fields. In addition, we ran deflated
GMRES-DR(40,10)\cite{morgan1} (the Krylov subspace is 40, the
number of deflated eigenvectors is 10.), which we find to be
adequate for our needs, and GMRES-Proj\cite{morgan2} to use the
deflated eigenvectors in solving other right hand sides. We have
studied in particular the zero momentum scalar matrix element, $S
= <\bar{\psi}\psi>$, in the presence of a nucleon. The quantity
that we form is
\begin{equation}
R_N(t,t')= \frac{G_N^3(t,t')}{G_N^2(t)},
\end{equation}
where $G_N^3(t,t')$ is the three-point function for the zero
momentum scalar insertion, and $G_N^2(t)$ is the two-point
function for the unpolarized nucleon. The three point function is
formed as
\begin{equation}
G_N^3(t,t') = <G_N^2(t)S(t')> - <G_N^2(t)><S(t')>,
\end{equation}
where $<\dots>$ denotes a gauge field average and $S(t')$ is the
raw disconnected scalar loop (we have not applied a
renormalization factor). We compute the quantity,
\begin{equation}
R_N(t) = \sum_{t'=source}^{t} R_N(t,t') \longrightarrow const +
G_S^{dis}(0), \label{signal}
\end{equation}
which gives the disconnected part of the scalar form factor,
$G_S^{dis}(0)$, at zero momentum. Our results (\lq\lq analysis 1")
for this quantity on 135 configurations with a single noise per
configuration are shown in Fig.~5. We use a zero-momentum nucleon
wall source at time step 4 for the two point function. We do not
attempt perturbative subtraction because of our finding, in the
previous section, that this helps little. The quantity $R_N(t)$ is
plotted as a function of the final nucleon and scalar loop
position, $t$. In addition, we show another analysis (\lq\lq
analysis 2") in which the upper limit in Eq.(\ref{signal}) is not
$t$, but fixed at 20 background time steps. We are looking for a
linear signal in $t$; the first analysis yields smaller error
bars. The value yielded by the 10-13 time step fit is
$G_S^{dis}(0) \approx 1.1\pm 1.3$. Although our error bar is
large, the sign and magnitude of our result is consistent with
expectations\cite{two}.

\section{Acknowledgements}

This work was supported in part by the National Science Foundation
under grant 0310573 and by the Natural Sciences and Engineering
Research Council of Canada. The calculations were done on the high
performance cluster at Baylor University.

\end{document}